\documentclass[letter]{aa}
\usepackage[varg]{txfonts}
\usepackage{booktabs,lipsum}
\usepackage{natbib}
\usepackage{rotating}
\usepackage{pdflscape}
\usepackage{color}
\usepackage{epsf} 
\usepackage{graphicx,epstopdf}

\def \ha {H$\alpha$}
\def \hb {H$\beta$}

\newcommand{\mdot}{M$_{\odot}$}

\newcommand{\changed}[1]{{\normalfont #1}}

\begin{document}

  \title{The Close AGN Reference Survey (CARS)} 

\subtitle{Mrk 1018 returns to the shadows after 30 years as a Seyfert 1}

\author{ R. E. McElroy \inst{1,2}, 
                B. Husemann $^{3,\star}$, 
        S. M. Croom \inst{1,2}, 
        T. A. Davis \inst{4}, 
        V. N. Bennert \inst{5}, 
        G. Busch \inst{6}, 
        F. Combes \inst{7}, 
        A. Eckart\inst{6,8}, 
        M. Perez-Torres \inst{9}, 
        M. Powell \inst{10}, 
        J. Scharw\"achter  \inst{11}, 
                G. R. Tremblay$^{10,\dagger}$
                T. Urrutia \inst{12}}

\institute{ Sydney Institute for Astronomy, School of Physics, University of Sydney, NSW 2006, Australia  \\
                        \email{rmcelroy@physics.usyd.edu.au}
                        \and ARC Centre of Excellence for All-sky Astrophysics (CAASTRO) 
            \and European Southern Observatory (ESO), Karl-Schwarzschild-Str. 2, D-85478 Garching b. M\"unchen, Germany, $\star$ ESO Fellow
            \and School of Physics \&\ Astronomy, Cardiff University, Queens Buildings, The Parade, Cardiff, CF24 3AA, UK
            \and Physics Department, California Polytechnic State University, San Luis Obispo, CA 93407, USA            
            \and I. Physikalisches Institut, Universit\"at zu K\"oln, Z\"ulpicher Stra\ss e 77, 50937 K\"oln, Germany
            \and LERMA, Observatoire de Paris, College de France, PSL, CNRS, Sorbonne Univ., UPMC, F-75014 Paris, France
            \and Max-Planck-Institut f\"ur Radioastronomie, Auf dem H\"ugel 69, D-53121 Bonn, Germany             
            \and Instituto de Astrophisica de Andalucia (IAA),  Glorieta de la Astronomía, s/n, ES-18008 Granada, Spanien
            \and Yale Center for Astronomy and Astrophysics, Yale University, 52 Hillhouse Ave., New Haven, CT 06511, USA, $\dagger$ Einstein Fellow
            \and Gemini Observatory, Northern Operations Center, 670 N. A'ohoku Place, Hilo, HI, 96720, USA
            \and Leibniz-Institut f\"uer Astrophysik Potsdam (AIP), An der Sternwarte 16, D-14482 Potsdam, Germany}

\titlerunning{Mrk 1018's return to the shadows}
\authorrunning{R. E. McElroy, B. Husemann, S. M. Croom\inst{1,2}, et al. }

\date{Draft as of \today}

\abstract{
We report the discovery that the known `changing look' AGN Mrk~1018 has changed spectral type for a second time. New VLT-MUSE data taken in 2015 as part of the Close AGN Reference Survey (CARS) shows
that the AGN has returned to its original Seyfert 1.9 classification. The CARS sample is selected to contain only bright type 1 AGN, but Mrk~1018's broad emission lines and continuum, typical of 
type 1 AGN, have almost entirely disappeared. We use spectral fitting of the MUSE spectrum and previously 
available spectra to determine the drop in broad line flux and the Balmer decrement. We find that the broad line flux has decreased by a factor of  $4.75 \pm 0.5$ in \ha\ 
since an SDSS spectrum was taken in 2000.  The Balmer decrement  has not changed significantly implying no enhanced
reddening with time, but the remaining broad lines are more asymmetric than those present in the type 1 phase. We posit that the change is due to an intrinsic drop in flux from the 
accretion disk rather than variable extinction or a tidal disruption event.
} 
\maketitle

\section[]{Introduction}
The various spectral classes of active galactic nuclei (AGN) have been unified based on the inclination of the accreting black hole (BH) to the line of sight \citep[see][]{Antonucci1993}. Under this 
scheme type 1 AGN, distinguished by their extremely broad permitted optical emission lines, are inclined to the observer such that we have a direct view onto the accretion disk. Type 2 AGN, devoid of 
these broad emission lines, possess only narrow lines suggesting that radiation from their accretion disks is blocked by what is commonly assumed to be a dusty torus. 
Intermediate objects complicate this simple dichotomy and are labelled as type 1.2, 1.5, 1.8, or 1.9 depending on the breadth and relative flux of narrow and broad emission lines 
\citep{Osterbrock1976}. 

\begin{figure*}
\centering
\includegraphics[width=0.895\textwidth]{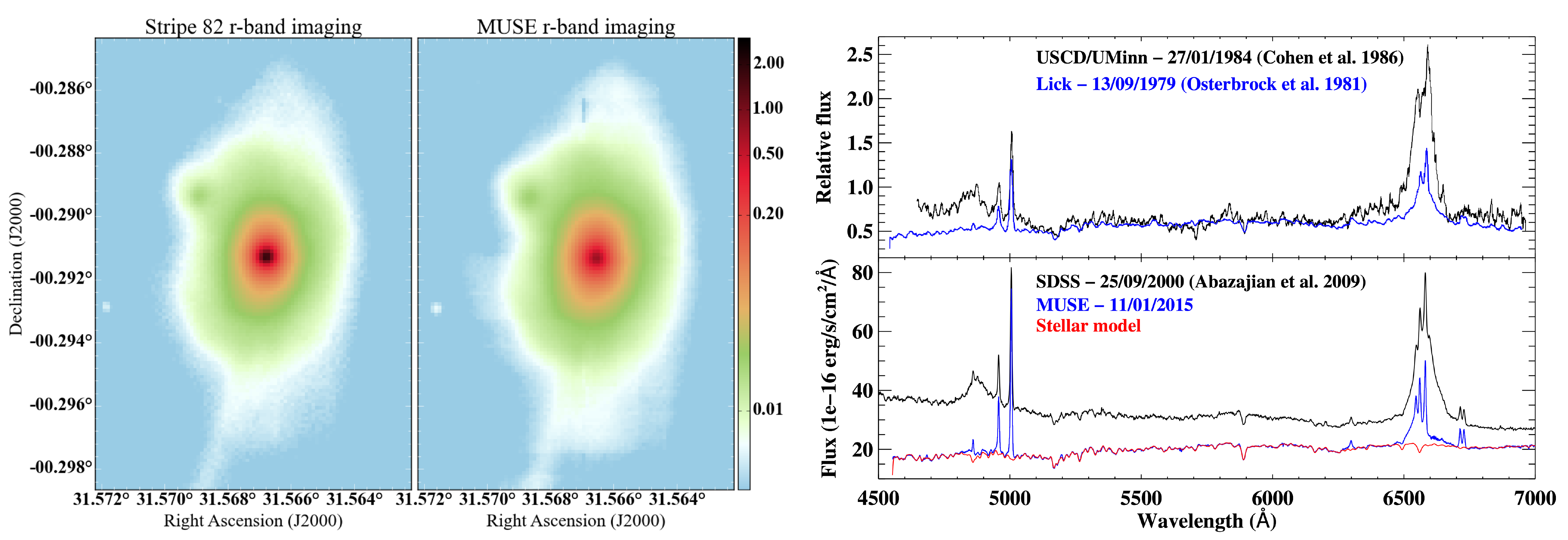}
\caption{{\it Left:} Comparison of the stacked Stripe 82 and re-constructed MUSE $r$ band image. The images are scaled logarithmically in units of $10^{-16} $erg/s/cm$^2/\AA$.  
{\it Right:} The top panel shows the original type 1.9 spectrum from 1979 \citep{Osterbrock1981} and the follow-up type 1 spectrum from 1984 \citep{Cohen1986} in units of relative flux. These spectra 
were extracted from the PDFs of the aforementioned papers, and scaled such that their continuum levels and narrow emission lines matched. The bottom panel shows the SDSS spectrum (2000) and a  
3" aperture spectrum from the MUSE data cube (2015), with the fit to the stellar component of the galaxy overplotted in red. }
\label{image}
\end{figure*}

Mrk 1018, a late-stage merger at $z = 0.043$ (see Figure \ref{image}), was one of the earliest examples of a `changing look' AGN, changing from a type 1.9 to 1 over the course of $<$ 5 years 
\citep{Cohen1986}. Since then a number of galaxies have 
been found to change type optically \citep[e.g.][]{Aretxaga1999, Denney2014} or in the X-ray domain \citep[e.g.][]{LaMassa2015, Ricci2016}. A much smaller sample of AGN have changed 
spectral type not once, but twice. Only a few galaxies have been observed to undergo this full cycle. Mrk 590 was observed to change from a type 1.5 to 1.0 from 1973 to 1989, then to gradually lose 
all evidence of broad lines until 2014 when it appeared as a type 2 AGN \citep{Osterbrock1977, Denney2014}. Similarly NGC 4151, originally classified as a Seyfert 1.5, lost then regained its broad 
emission lines \citep{Osterbrock1977, Antonucci1983, Shapovalova2010}. 

Recent work on accretion timescales suggests that AGN flicker on short timescales, $\sim$\,$10^5$ years, rather than remaining at a constant 
luminosity \citep{Schawinski2015}. \cite{Elitzur2014} posit that AGN may not be statically type 1 or type 2 throughout their lifetimes, and can change type as they increase or decrease 
in luminosity. AGN that have changed type can help us to understand the accretion physics and associated timescales. 

A few hypotheses are used to explain why AGN change spectral type. Variable obscuration of the nuclear region or a patchy torus would allow for the same object to be viewed as a type 1 
or type 2 AGN over time \citep{Elitzur2012}. In this scenario, the intervening absorption along our line of sight is changing and not the intrinsic luminosity. 
An alternative option is that a drop (or increase) in the accretion rate onto the BH leads to a different AGN classification. \cite{Merloni2015} discuss how tidal disruption 
events (TDE) of stars close the central BH cause AGN to flare and dramatically increase their luminosity leading to their classification as changing look AGN. 

In this paper, we present the serendipitous discovery that Mrk 1018 has changed type again. New data show that Mrk 1018 presents spectral signatures typical of a type 1.9 AGN once more, 
with drastically dimmed broad lines and continuum. 

\section[]{Observations and data analysis}
\subsection[]{MUSE 3D spectroscopy}

\begin{figure*}
\centering
\includegraphics[width=0.9\textwidth]{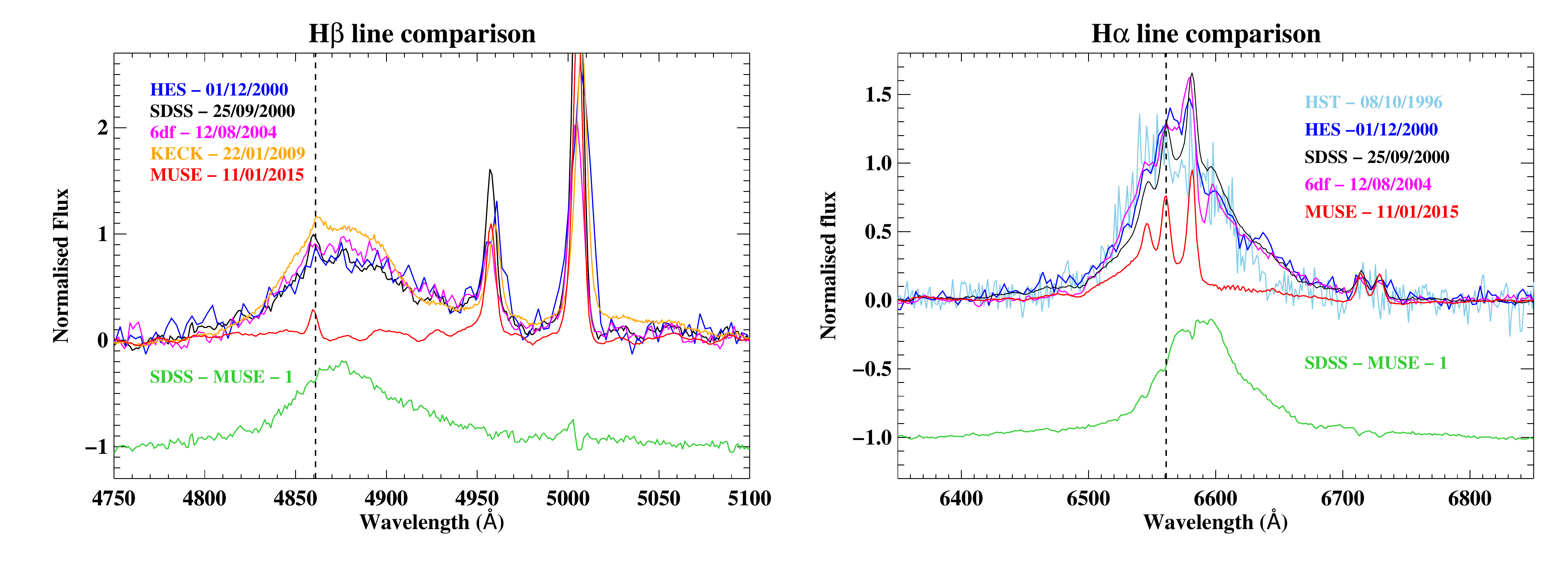}
\caption{Archival spectra focussed on  the \hb\ (left panel) and \ha\ (right panel) emission lines. The spectra are plotted in continuum subtracted normalised flux to account for any discrepancies in 
flux calibration, units, or aperture size. 
The systemic redshift of \ha\ and \hb\ based on the narrow emission lines is 
indicated by the dashed black line. The Keck spectrum from 2009 is plotted in orange and covers only the \hb\ line. Residuals of the SDSS spectrum (2000) minus our MUSE 
spectrum (2014) are plotted in green to emphasise the change in flux and line shape.}
\label{spectra}
\end{figure*}

We observed Mrk~1018 with the Multi-Unit Spectroscopic Explorer \citep[MUSE,][]{Bacon2010} at the Very Large Telescope on 2015 Jan 11 as part of the Close AGN Reference Survey (CARS, 
\url{cars-survey.org}). CARS aims to provide a detailed multiwavelength view of 40 nearby ($0.01 < z < 0.06$) type 1 AGN. 

Mrk~1018 was observed under clear sky conditions at an average seeing of $1\farcs2$ and split into two 400\,s exposure with a rotation of the FoV by $90\degr$. We reduced the data with the ESO-MUSE 
pipeline v1.2.1 \citep{Weilbacher2012} using the standard calibrations. The sky background was estimated and subtracted from the object free signal in the large $1\arcmin\times1\arcmin$ field of view 
(FoV) of MUSE using the pipeline. The final cubes are sampled at $0\farcs2$ per spaxel and cover a wavelength range of 4750--9300$\AA$ with a spectral resolution of $2.4\AA$ (FWHM).

\begin{table}
  \centering
    \caption{Available spectra for Mrk~1018.}
    \begin{tabular}{lccc}
    \hline\hline
    Source & Date & Rest $\lambda$-range (\AA) & Ref. \\ 
    \hline
    Lick & 1979 Sept 13  &  4500 - 8800 & (1)  \\
    USCD/UMinn  & 1984 Jan 27  &  4500 - 6900 & (2)   \\
    HST  & 1996 Oct 8  &  6200 - 9200    \\
    SDSS  & 2000 Sept 25  &  3600 - 8800  & (3)  \\
    HES  & 2000 Dec 1  &  3400 - 8700   & (4) \\
    6df  & 2004 Aug 15  &  3700 - 7200  & (5)  \\
    Keck   & 2009 Jan 22  &  3400 - 5300  & (6) \\
    MUSE   & 2015 Jan 11  &  4500 - 9000  & (7) \\

    \hline
    \end{tabular}%
    \tablebib{(1)~\cite{Osterbrock1981}; (2)~\cite{Cohen1986}; (3)~\cite{Abazajian2009}; (4)~\cite{Wisotzki2000};
(5)~\cite{Jones2009}; (6)~\cite{Bennert2011}; (7) This work.}
  \label{spectra_table}%
\end{table}%

Mrk~1018 shows spectral features of a type 1.9 AGN (see Fig.~\ref{image}) with a nearly undetectable broad H$\beta$ line. However, several previous spectroscopic 
observations (listed in Table~\ref{spectra_table}) are consistent with a type 1 AGN. Thus, our new MUSE observations show that Mrk~1018 has changed optical spectral type again after 
$\sim$30 years of being a luminous unobscured AGN \footnote{ \changed{Continuous photometry and spectroscopy are not available for the entire 30 years, but the FUV observations from 1984 and 1996 are 
consistent in flux implying little change during that period.}}.

\subsection[]{Continuum and emission line analysis}

We take the SDSS spectrum (Table \ref{spectra_table}) to be representative of Mrk~1018's bright state. We extract an equivalent MUSE spectrum using a 3" diameter aperture centred on the AGN position and correct 
for absolute flux calibration offsets by matching the strength of stellar absorption lines. The MUSE spectrum has weaker AGN continuum and broad emission lines allowing us to model the stellar 
continuum by fitting a super-position of Indo-U.S. Library of Coud\'e Feed Stellar Spectra \citep{Valdes2004} with {\sc ppxf} \citep{Cappellari2004} excluding regions with emission lines. The MUSE 
spectrum and the best-fit continuum model are shown on the bottom right of Fig.~\ref{image}. We subtract this stellar contribution from the MUSE \textit{and} SDSS spectra.  

A faint broad \hb\ emission line is clearly recovered in the MUSE data after continuum subtraction. We fit both the narrow and broad emission lines as a super-position of Gaussians and an additional 
local power-law continuum for the SDSS spectrum. The light travel time to the narrow line region (NLR) is likely hundreds of years. As such, no significant change in flux is expected within ten years, which 
is consistent with the data. For accurate \ha\ and \hb\ emission line fluxes we must rely on the extrapolated stellar continuum model, which is made up of a subset of the stellar templates. Uncertainty due to template mismatch is simulated by selecting a random subset of 16 templates from the 504 in the full Indo-U.S. library and repeating the fitting 100 times. We then measure the variance of the output fluxes. 

From the emission-line modelling we obtain broad \ha\ line fluxes of $(3.3\pm 0.1)\times10^{-13}\,\mathrm{erg}\,\mathrm{s}^{-1}\,\mathrm{cm}^{-2}$ and 
$(7.0\pm 0.2 )\times10^{-14}\,\mathrm{erg}\,\mathrm{s}^{-1}\,\mathrm{cm}^{-2}$ for the SDSS and MUSE spectra, respectively, corresponding to a flux decrease of a factor of 
$4.79 \pm 0.3$. \changed{Here we adopt $H_0=70\,\mathrm{km}\,\mathrm{s}^{-1}\,\mathrm{Mpc}^{-1}$, $\Omega_{\mathrm{m}}=0.3$, and $\Omega_\Lambda=0.7$}. Similarly, we obtain broad \hb\ line fluxes of $(8.0\pm 
0.1)\times10^{-14}\,\mathrm{erg}\,\mathrm{s}^{-1}\,\mathrm{cm}^{-2}$ and $(2.2\pm 0.2)\times10^{-14}\,\mathrm{erg}\,\mathrm{s}^{-1}\,\mathrm{cm}^{-2}$ corresponding to \hb\ dimming of a factor of 
$3.57 \pm 0.5$. The Balmer decrement of \ha\ to \hb\ is $3.3 \pm 0.8$ for the MUSE spectrum and $4.2\pm 0.4$ for the SDSS spectrum. Extinction has not increased as new obscuration along the line of 
sight to the BLR would lead to an increased Balmer decrement.

The evolution of the \hb\ and \ha\ emission line profiles are shown in Fig.~\ref{spectra}  over the period 1996-2014. While the broad-line shape remains almost unchanged during the bright 
state, our most recent MUSE spectrum shows a drastically different profile. More flux is lost on the red wing, \changed{as shown by the 
difference between the SDSS and MUSE spectra (green lines) in Fig.~\ref{spectra}}. The residual is redshifted with respect to the systemic redshift.
The Keck spectrum was the last one obtained during the bright state and the \hb\ line shows a decrease on the red 
side. We find broad \ha\ line width (FWHM) of $4000\pm 100\,\mathrm{km\,s}^{-1}$ (SDSS) and $3300\pm 200\,\mathrm{km\,s}^{-1}$ (MUSE). 
We expect the broad line width to increase as the AGN luminosity falls to preserve derived M$_{BH}$ estimates \citep{Ruan2015}. \changed{ However, following \cite{Woo2015} we infer single-epoch BH 
masses of $\log(M_\mathrm{BH}/M_\sun)=7.9$ and $\log(M_\mathrm{BH}/M_\sun)=7.4$ in the high and low states, respectively.} 
\changed{This} may suggest that the BLR is not in an equilibrium state or the virial factor has changed.

\begin{figure}
\centering
\includegraphics[width=0.48\textwidth]{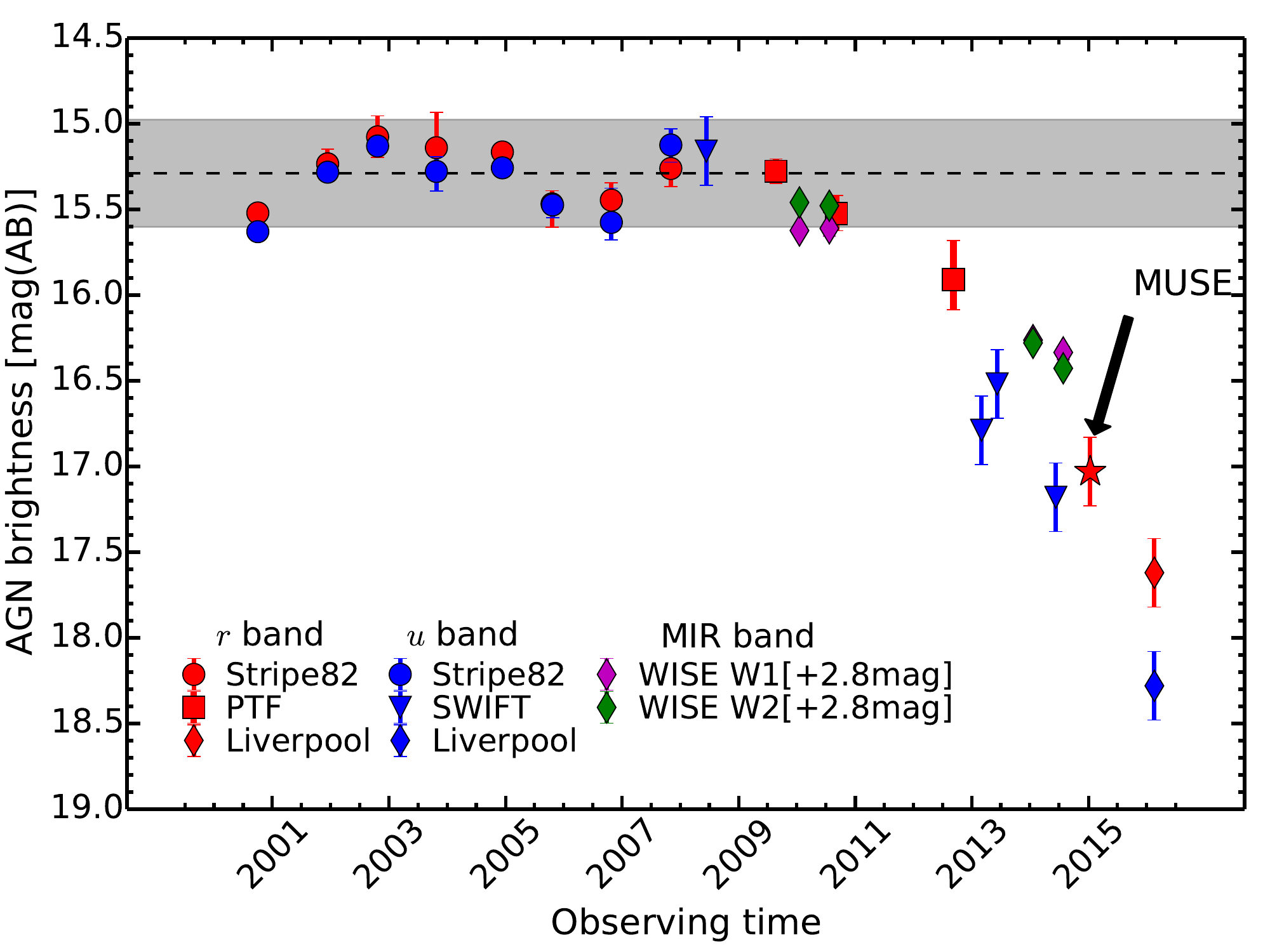}
\caption{Time series of archival photometric data for Mrk 1018. Optical $r$ band data is shown in red, $u$ band in blue, and 
MIR in purple and green. The dashed line shows the mean magnitude of the core (15.28 mag) in the r band and the grey band is the $\pm$2-sigma scatter. The photometry indicates a rapid decline in AGN 
flux beginning around 2011. }
\label{photometry}
\end{figure}

\subsection[]{Photometry}
We study the long-term variability of Mrk~1018 with archival observations in MIR and optical ($u$ and $r$) bands. These include SDSS Stripe 82 
\citep{Abazajian2009}, Palomar transient factory  \citep[PTF,][]{Law2009}, \textit{SWIFT} UV-optical telescope \citep{Roming2005}, Liverpool telescope imaging (February 2016) in the optical, and \text{WISE} in the MIR \citep{Wright2010}. 

It is crucial to decompose the AGN and host galaxy light to infer the intrinsic variability of the nucleus. We estimate the 2D surface brightness 
distribution of the host galaxy by fitting a simple two-component -- PSF plus Sersi\'c -- to the deep $u$ and $r$ band co-added Stripe 82 images with GALFIT \citep{Peng2002}. The best-fit 
parameters for the Sersi\'c model are slightly different with $r_\mathrm{e}=14\farcs6$, $n=2.23$ and $b/a=0.56$ for the $u$ band and $r_\mathrm{e}=9\farcs9$, $n=2.77$, and $b/a=0.62$ for the $r$ band. 
We then fit each individual image keeping those parameters fixed for the $u$ and $r$ band, and use an empirically constrained PSF from 
a bright unsaturated star close to Mrk~1018. Only integrated magnitudes can be reported for the \textit{WISE} images.

The late evolution in the AGN light curve is shown in Fig.~\ref{photometry}. The mean magnitude of the \changed{nucleus} was 15.28 mag (AB) in the r band with a scatter of 0.15\,dex (1$\sigma$) prior 
to the rapid decline. \changed{The constant host magnitude within 5” is 15.57 mag (AB), implying that the luminosity of the AGN has dropped from 1.3 to 0.3 times that of the host.}
The latest data point from Liverpool telescope indicates that the AGN brightness is still declining.

 \section[]{Discussion}

Our spectroscopic analysis shows that the \ha\ broad emission line flux in the MUSE spectrum (2015) is a factor of $\sim$4.75 lower compared to the SDSS spectrum (2000) and has 
changed shape markedly. The multi-epoch photometry shown in Fig.~\ref{photometry} indicates that rather than peaking in luminosity before dimming to its current state, Mrk 1018 had a relatively
constant luminosity before dropping. Below we discuss the possible scenarios that could have led to the apparent dimming of the nucleus.

The tidal disruption of a massive star by the central BH can cause AGN to change spectral type due to a flare in accretion. A tidally disrupted star temporally provides fuel for the nucleus until it is destroyed or is too far away \citep{Merloni2015}. We calculate the mass accreted during the bright phase assuming that the luminosity remained 
constant. Our measurements of $L_{\mathrm{H}\alpha}$ and M$_\mathrm{BH}$ from the SDSS spectrum lead to a bolometric luminosity of  $L_{\rm{bol}} = 3.1 \times  10^{44} \mathrm{erg\,s}^{-1}$ and 
Eddington ratio of $L_\mathrm{bol}/L_\mathrm{Edd} = 0.03$.
Taking the efficiency of an accretion disk to be 0.1, this gives a mass accretion rate of 0.055 \mdot$\,\mathrm{yr}^{-1}$ or a total mass accreted over the bright phase ($\approx$ 30 
years) of 1.65 \mdot. Typically, $\sim$ 50\% of the disrupted star's mass is accreted, requiring a 3.2 \mdot\ star. While this could come from a single star, 
a TDE is characterised by a rapid increase in AGN luminosity, which then peaks before decreasing again over a maximum of several years \citep{Guillochon2013}. We see a 30-year plateau in Mrk 1018's 
 luminosity and no defined peak (only the last 10 years are shown in Fig. \ref{photometry}). To our knowledge, no TDE model predicts a bright phase this prolonged, which leads us to conclude 
that it is unlikely that a TDE caused Mrk 1018's recent change.

Could Mrk 1018's change be the result of a declining accretion rate?  
As the ionising radiation from an AGN falls, the broad line emission comes from faster moving BLR gas closer to the BH. Observationally, the BLR luminosity should drop, velocity dispersion should increase, and the shape of the lines may change. Therefore, if the altered broad line shape is due to decreased AGN luminosity, the velocity dispersion of the emission lines should increase. We observe the opposite.
The BLR represents a potential reservoir of fuel for the accretion disk approximately 24 light days away (following \cite{Bentz2013} using $L_{\mathrm{H}\alpha}$ converted to $L_{5100}$ \citep{Woo2015}).
This is equivalent to a radial inflow time of $\sim$ 5 years based on the measured velocity dispersion. If the declining accretion rate were due to gradual depletion of fuel we would not expect continuing BLR emission. A disruption of the accretion flow could cause the disc to be temporarily starved, while leaving the BLR unconsumed. The observed decrease -- rather than increase -- in 
velocity dispersion as the nuclear luminosity fell may be due to a change in the virial factor.
This would imply that the BLR structure and kinematics have been altered. Since Mrk 1018 
is a late-stage major merger, two SMBHs may orbit one another at the centre of the galaxy. Such a nearby massive object may cause instabilities in the accretion disk and BLR, leading to a 
rapid drop in accretion rate and changed emission line shape. We lack the spatial resolution to resolve a hypothetical dual SMBH system, which can only be done 
with long baseline radio interferometry.

We see no evidence for increased extinction, but it is possible that there is new gas or dust along our line of sight blocking our view to the AGN. To explain the lack of reddening this obstacle would 
have to be extremely opaque with a very high $A_{\rm{V}}$. Additionally, the obscuring material would have to be placed such that select parts of the BLR are still visible. This could explain the new line shape as the redshifted wing of the broad line may be in the region that is blocked from view. However, this argument requires a complex and convoluted 
geometry in order to explain our observations.

\section{Conclusions}

In this letter we presented evidence for the AGN Mrk~1018 returning
to its Seyfert 1.9 state after $\sim$30 years as a Seyfert 1. Using MUSE data from the Close AGN Reference Survey alongside archival spectroscopic and photometric observations we explored the possible causes of this change: a decline in accretion rate or lack of fuel, a TDE, obscuration, or disruption of the accretion disk. We reason that the length and consistency of Mrk~1018's bright phase makes a TDE an unlikely explanation, but we cannot rule out a simple decline in accretion rate. The Balmer decrement between the Seyfert 1 and 1.9 phases implies that the obscuration along our line of sight has not increased, but a highly opaque column of gas selectively obscuring our view is still a possibility.  

In light of this exciting discovery we were awarded Chandra, HST, and VLA director's discretionary time to investigate how Mrk 1018 has changed since our MUSE observations. While the data presented in 
this paper cannot definitively tell us the nature of the change in Mrk 1018's nucleus, these newer observations in the UV, X-ray, and radio bands will provide further constraints. 

\begin{acknowledgements}
 Based on observations made with ESO Telescopes at the La Silla Paranal Observatory under programme ID 94.B-0345 and the Liverpool Telescope operated on the island of La Palma by Liverpool John Moores 
University in the Spanish Observatorio del Roque de los Muchachos of the Instituto de Astrofisica de Canarias. 
Parts of this research were conducted by the Australian Research Council Centre of Excellence for All-sky Astrophysics, through project number CE110001020.
GRT acknowledges support from  NASA through Einstein Postdoctoral Fellowship Award Number PF-150128, issued by the Chandra X-ray Observatory 
Center, which is operated by the Smithsonian Astrophysical Observatory for and on behalf of NASA under contract NAS8-03060.
TAD acknowledges support from a Science and Technology Facilities Council Ernest Rutherford Fellowship.
VNB gratefully acknowledges assistance from a NSF Research at Undergraduate Institutions grant AST-1312296.  Findings and conclusions do not necessarily represent the views of the NSF.
MAPT acknowledges support from the Spanish MINECO through grant AYA2015-63939-C2-1-P.
\end{acknowledgements}

\bibliography{bib_mrk1018_v1}
\bibliographystyle{aa}

\end{document}